\documentclass[12pt,leqno]{amsart} 

\usepackage[utf8]{inputenc} 

\usepackage{algorithm}
\usepackage[noend]{algpseudocode}
\makeatletter
\def\BState{\State\hskip-\ALG@thistlm}
\makeatother

\usepackage{graphicx}
\usepackage{hyperref}
\hypersetup{
    unicode=false,          
    pdftitle={DV Curvature Transformation},    
    pdfkeywords={curvature}, 
    colorlinks=true,       
    pdfborder=0 0 0,
    linkcolor=blue,          
    citecolor=blue,        
    filecolor=blue,      
    urlcolor=blue           
}

\usepackage{geometry} 
\usepackage{graphicx} 
\usepackage{xcolor}

\usepackage{booktabs} 
\usepackage{array} 
\usepackage{paralist} 
\usepackage{verbatim} 
\usepackage{subfig} 
\usepackage[font=small,labelfont=small]{caption}

\usepackage{fancyhdr} 
\newcommand{\R}{\mathbf{R}}

\newcommand{\area}{\operatorname{area}}

\title{Curvature transformation}
\author{Dimitris Vartziotis}
\address{TWT GmbH Science \& Innovation, Department for Mathematical Research, Ernsthaldenstr. 17, 70565 Stuttgart, Germany }
\address{NIKI Ltd. Digital Engineering, Research Center, 205 Ethnikis Antistasis Street,
45500 Katsika, Ioannina, Greece$$ $$}
\email{dimitris.vartziotis@twt-gmbh.de }
\begin{document}
\maketitle
\thispagestyle{empty}

\begin{abstract}
 A transformation based on mean curvature is introduced which morphs triangulated surfaces into round spheres.
\end{abstract}

\section{Introduction}

Let $S \subset \mathbb{R}^3$ be a surface embedded in $\mathbb{R}^3$ which is locally defined by the smooth parametrization $\left\{(x,y,S(x,y))\;\big|\; (x,y) \in U \subset \mathbb{R}^2\right\} \subset S$. 
The mean curvature $H(x,y)$ at every point $z=S(x,y)$ is given as the mean of the maximal and minimal principal curvatures: 
$$H(x,y)=\frac{1}{2}(\kappa_1 + \kappa_2).$$
The sphere is known to be the unique embedded constant mean curvature (CMC) surface \cite{aleksandrov1958}, and the unique immersed CMC sphere \cite{hopf1983}. For higher genus, there are many constant mean curvature surface. 
The vector mean curvature given by $\nu H(x,y)$ where $\nu$ is the outer unit normal vector at $z=S(x,y)$ is the negative gradient of the area functional of $S$, that is
$$\nu H=-\nabla\area(S).$$
The parametric family $\left\{S(x,y,t)\right\}_{t\in \mathbb{R}}$ of surfaces which evolves proportionally to the vector mean curvature, so that the area of $S(t)$ reduces with $ t \rightarrow \infty$, is called \emph{mean curvature flow} and is the solution of the following partial differential equation
\begin{equation}\label{e.meancurvature}
\frac{\partial S(x,y,t)}{\partial t} = - \nu H(x,y,t),
\end{equation}
where $H(x,y,t)$ is the mean curvature at the point $S(x,y,t)$ of the surface $S(t)$. Huisken \cite{huisken1984} proved that if $S(0)$ is a bounded convex surface, then $S(t)$ becomes more and more nearly spherical as it shrinks, and at the instant it vanishes it is asymptotic to the shrinking sphere given above.

It has always been our aim to transform and study polyhedra, triangulated surfaces and meshes in the context of mesh smoothing by elementary geometric means \cite{VartziotisAthanasiadisGoudasWipper2008,VartziotisWipperPapadrakakis2013,VartziotisBohnet2014,VartziotisHimpel2014}. This is the motivation behind studying mean curvature using elementary geometric transformations. Given a triangulated surface $S$, the discrete mean curvature is defined for each edge $e$ by
\[K(e) = l(e) \theta(e),\] where $\theta(e)\in (-\pi,\pi]$ is the (oriented) dihedral angle between the two adjacent facets and $l(e)$ is the length of the edge. If $n_1$ and $n_2$ are the (oriented) unit surface normals, then $\cos \theta(e) = n_1 \cdot n_2$. For each vertex $p$, its mean curvature $K(p)$ is defined by averaging over the mean curvatures of the neighboring edges.
We tried to find a simple transformation based on the discrete mean curvature which morphs a (reasonably shaped) triangulated surface into a round sphere.

\section{The transformation}\label{sec:transformation}

Let $S$ be an oriented, triangulated surface embedded in $\R^3$ given by a set of vertices $V \subset \R^3$ and a set of (unoriented) edges $E\subset V \times V$. Since $S$ is embedded in $\R^3$ there is an inside and and outside, and we choose unit face normals to point outward. The vertex normal $n_p$ at $p$ is computed by averaging the unit face normals of the adjacent faces and then normalizing the result. Let $K_\text{min}$ and $K_\text{max}$ be the minimal and maximal discrete mean curvature for the vertices of an oriented mesh. The transformation consists of applying two steps iteratively, each of which is applied to all points simultaneously.
\begin{align}
\label{move_in} I_C(p) & = p - C \frac{K(p) - K_\text{min}}{K_\text{max}-K_\text{min}} n_{p},\\
\label{move_out} O_C(p) & = p + C \left(1-\frac{K(p) - K_\text{min}}{K_\text{max}-K_\text{min}}\right) n_p.
\end{align}

Transformation $I_C$  in \eqref{move_in} moves each point inward, and $O_C$ in \eqref{move_out} moves each point outward. The factors in front of $n_p$ are chosen so that the magnitude of the translating vector is a value in the interval $[0,1]$, and it is big for small curvature values and vice versa. The dynamic process is determined by the following triple: $(k_\text{in},k_\text{out},C)$. The integers $k_\text{in}$ and $k_\text{out}$ determine how often the outward resp. inward transformation is applied before the other one is applied. More specifically we consider the transformation
\[
 T_{(k_\text{in},k_\text{out},C)} = O_C^{k_\text{out}} \circ I_C^{k_\text{in}}.
\]
This transformation $T$ will be repeated $n$ times as described in Algorithm \ref{alg:morph-step}.

\begin{algorithm}[h]
\caption{Curvature transformation}\label{alg:morph-step}
\begin{algorithmic}
\Function{Morph-Step}{$n,k_\text{in},k_\text{out},C,V$}
\For {$i\gets 1,\ldots,n$}
  \For {$j\gets 1,\ldots,|V|$}
\State    $p'_j \gets T_{(k_\text{in},k_\text{out},C)}(p_j)$
  \EndFor
  \For {$j\gets 1,\ldots,|V|$}
\State    $p_j \gets p'_j$
  \EndFor
\EndFor
\State {\bf return} $V$
\EndFunction
\end{algorithmic}
\end{algorithm}

We will see in Section \ref{sec:tests} that the constant factor $C = 0.25$ makes the transformation process described in Algorithm \ref{alg:spherify} converge towards a CMC sphere for a variety of surfaces. Even though this behavior is not surprising, it is difficult to find conditions that guarantee convergence and a proof thereof.

\begin{algorithm}
\caption{Curvature transformation}\label{alg:spherify}
\begin{algorithmic}[1]
\Procedure{Morph}{$m,V$}
\For {$i\gets 1,\ldots,m$}
  \State	$V \gets \text{Morph}(100,2,2,0,25,V)$
  \State	$V \gets \text{Morph}(100,2,1,0,25,V)$
\EndFor
\EndProcedure
\end{algorithmic}
\end{algorithm}

\section{Numerical tests}\label{sec:tests}

This transformation has been implemented in Python using the mean curvature method provided by the VTK class vtkCurvatures. We have tried different parameters and found that the following process works on our examples. The number of iterations mentioned in the figures corresponds to $200\cdot m$.

\bigskip
\bigskip
\bigskip

I want to thank George Philos from NIKI and Benjamin Himpel from TWT for their support in writing this research note.

\quad \quad D. Vartziotis

\bigskip
\bigskip

\begin{figure}
  \centering
  \begin{minipage}{0.4\textwidth}
  \includegraphics[width=\textwidth]{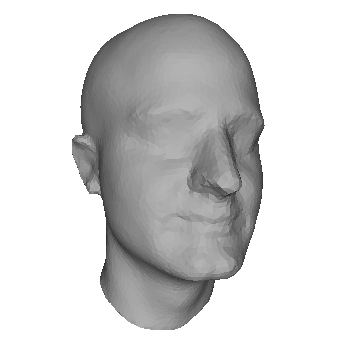}
  \end{minipage}
  \begin{minipage}{0.4\textwidth}
  \includegraphics[width=\textwidth]{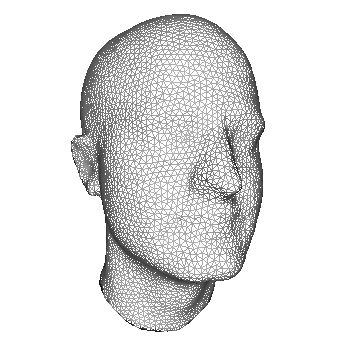}
  \end{minipage}
  \caption{Initial face mesh}
  \label{fig1}
\end{figure}

\begin{figure}
  \centering
  \begin{minipage}{0.4\textwidth}
  \includegraphics[width=\textwidth]{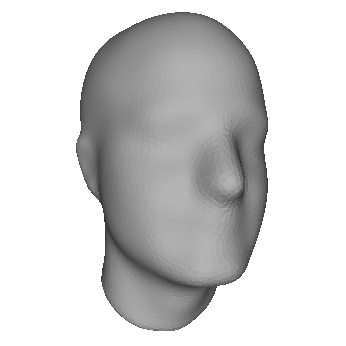}
  \end{minipage}
  \begin{minipage}{0.4\textwidth}
  \includegraphics[width=\textwidth]{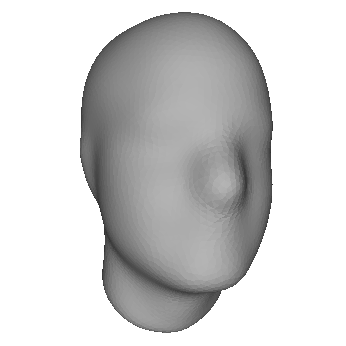}
  \end{minipage}
  \caption{Face mesh after 100 (left) and 200 (right) iterations}
  \label{fig2}
\end{figure}

\begin{figure}
  \centering
  \begin{minipage}{0.4\textwidth}
  \includegraphics[width=\textwidth]{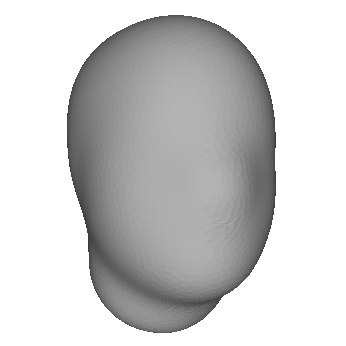}
  \end{minipage}
  \begin{minipage}{0.4\textwidth}
  \includegraphics[width=\textwidth]{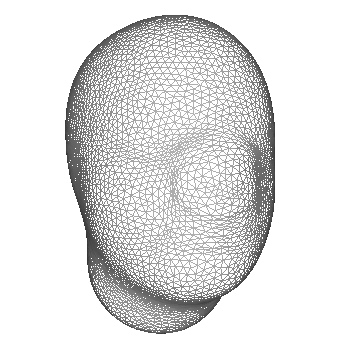}
  \end{minipage}
  \caption{Face mesh after 1000 iterations}
  \label{fig3}
\end{figure}

\begin{figure}
  \centering
  \begin{minipage}{0.4\textwidth}
  \includegraphics[width=\textwidth]{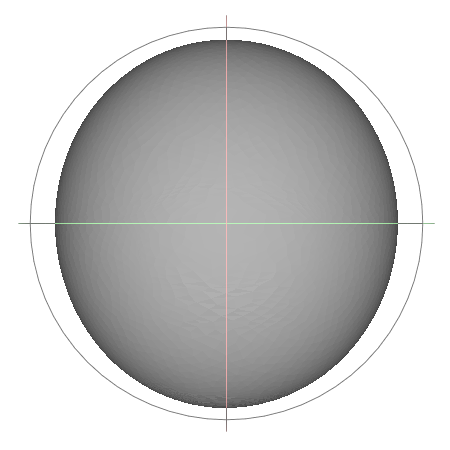}
  \end{minipage}
  \begin{minipage}{0.4\textwidth}
  \includegraphics[width=\textwidth]{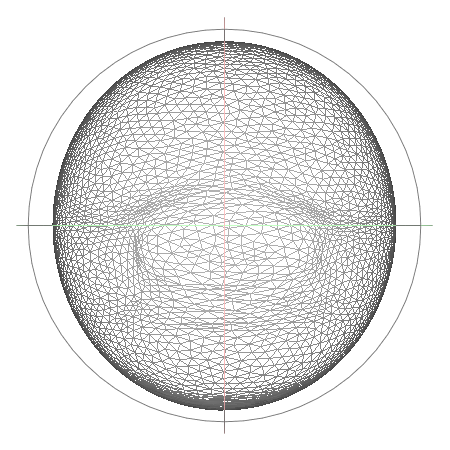}
  \end{minipage}
  \caption{Face mesh after 10000 iterations}
  \label{fig4}
\end{figure}

\begin{figure}
  \centering
  \begin{minipage}{0.4\textwidth}
  \includegraphics[width=\textwidth]{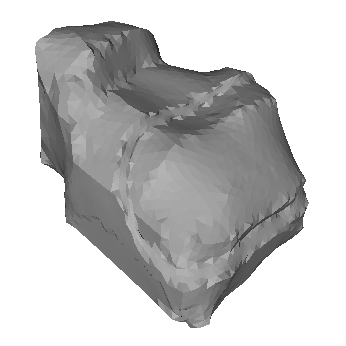}
  \end{minipage}
  \begin{minipage}{0.4\textwidth}
  \includegraphics[width=\textwidth]{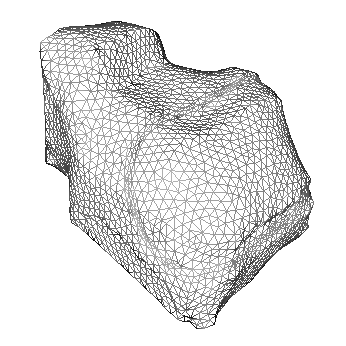}
  \end{minipage}
  \caption{Initial stone mesh}
  \label{fig5}
\end{figure}

\begin{figure}
  \centering
  \begin{minipage}{0.4\textwidth}
  \includegraphics[width=\textwidth]{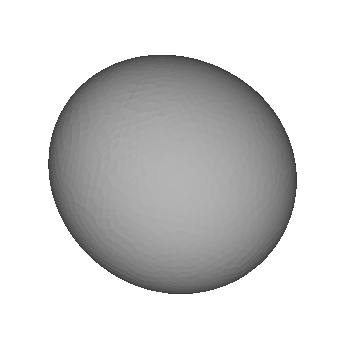}
  \end{minipage}
  \begin{minipage}{0.4\textwidth}
  \includegraphics[width=\textwidth]{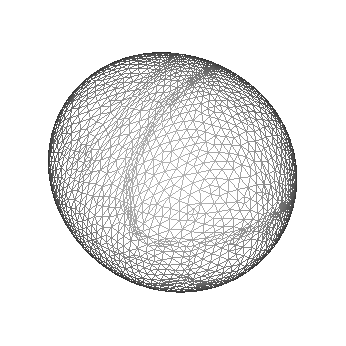}
  \end{minipage}
  \caption{Stone mesh after 1500 iterations}
  \label{fig6}
\end{figure}

\begin{figure}
  \centering
  \begin{minipage}{0.4\textwidth}
  \includegraphics[width=\textwidth]{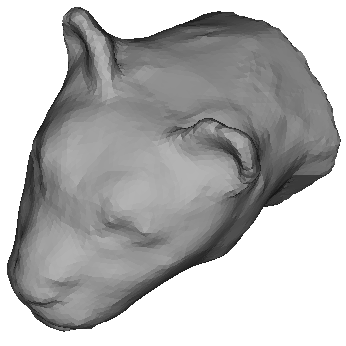}
  \end{minipage}
  \begin{minipage}{0.4\textwidth}
  \includegraphics[width=\textwidth]{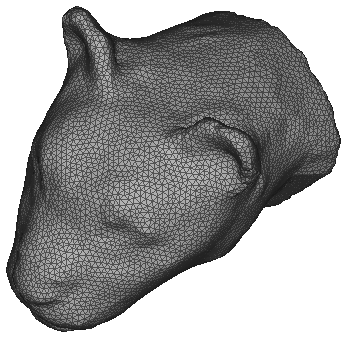}
  \end{minipage}
  \caption{Initial puma mesh}
  \label{fig7}
\end{figure}

\begin{figure}
  \centering
  \begin{minipage}{0.4\textwidth}
  \includegraphics[width=\textwidth]{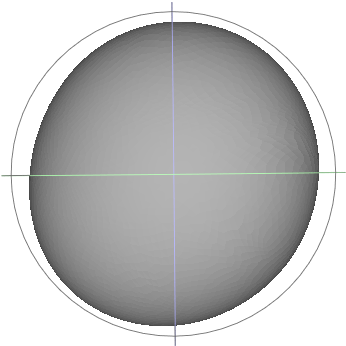}
  \end{minipage}
  \begin{minipage}{0.4\textwidth}
  \includegraphics[width=\textwidth]{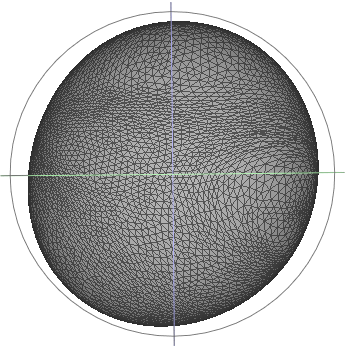}
  \end{minipage}
  \caption{Puma mesh after 2000 iterations}
  \label{fig8}
\end{figure}

\begin{figure}
  \centering
  \begin{minipage}{0.4\textwidth}
  \includegraphics[width=\textwidth]{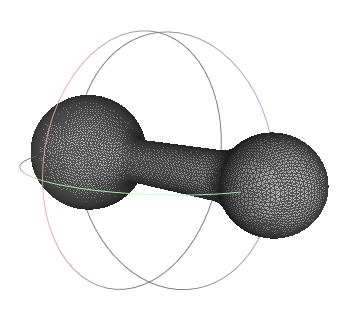}
  \end{minipage}
  \begin{minipage}{0.4\textwidth}
  \includegraphics[width=\textwidth]{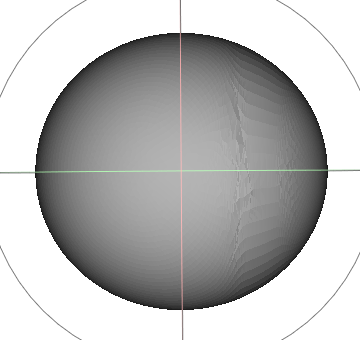}
  \end{minipage}
  \caption{Sphere bar mesh initial (left) and after 7500 iterations}
  \label{fig9}
\end{figure}

\begin{figure}
  \centering
  \includegraphics[width=0.7\textwidth]{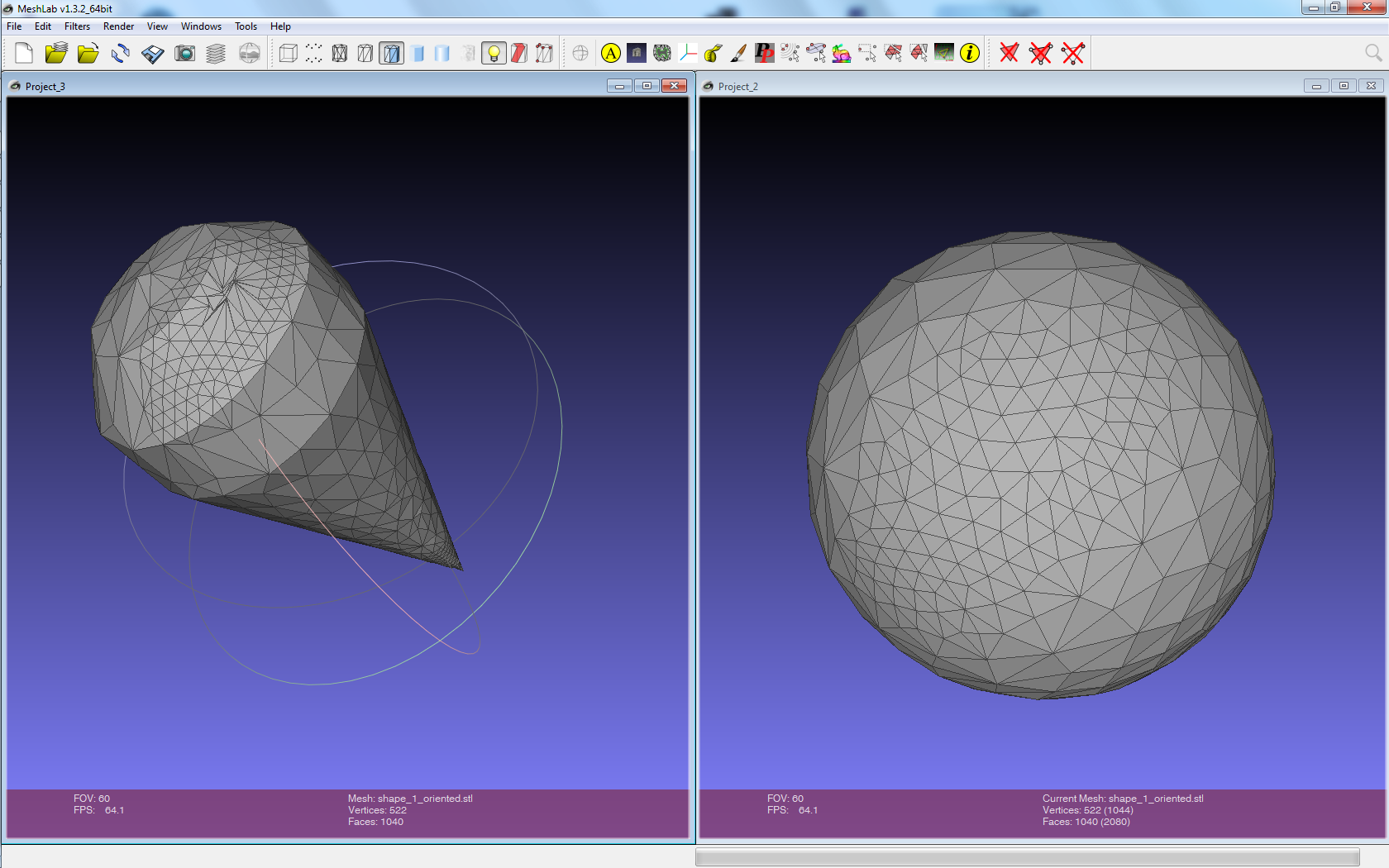}
  \caption{Initial mesh (left) and result (right)}
  \label{fig10}
\end{figure}

\begin{figure}
  \centering
  \includegraphics[width=0.8\textwidth]{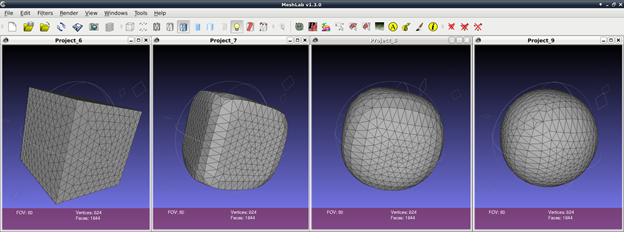}
  \caption{cube mesh}
  \label{fig11}
\end{figure}

\begin{figure}
  \centering
  \begin{minipage}{0.4\textwidth}
  \includegraphics[width=\textwidth]{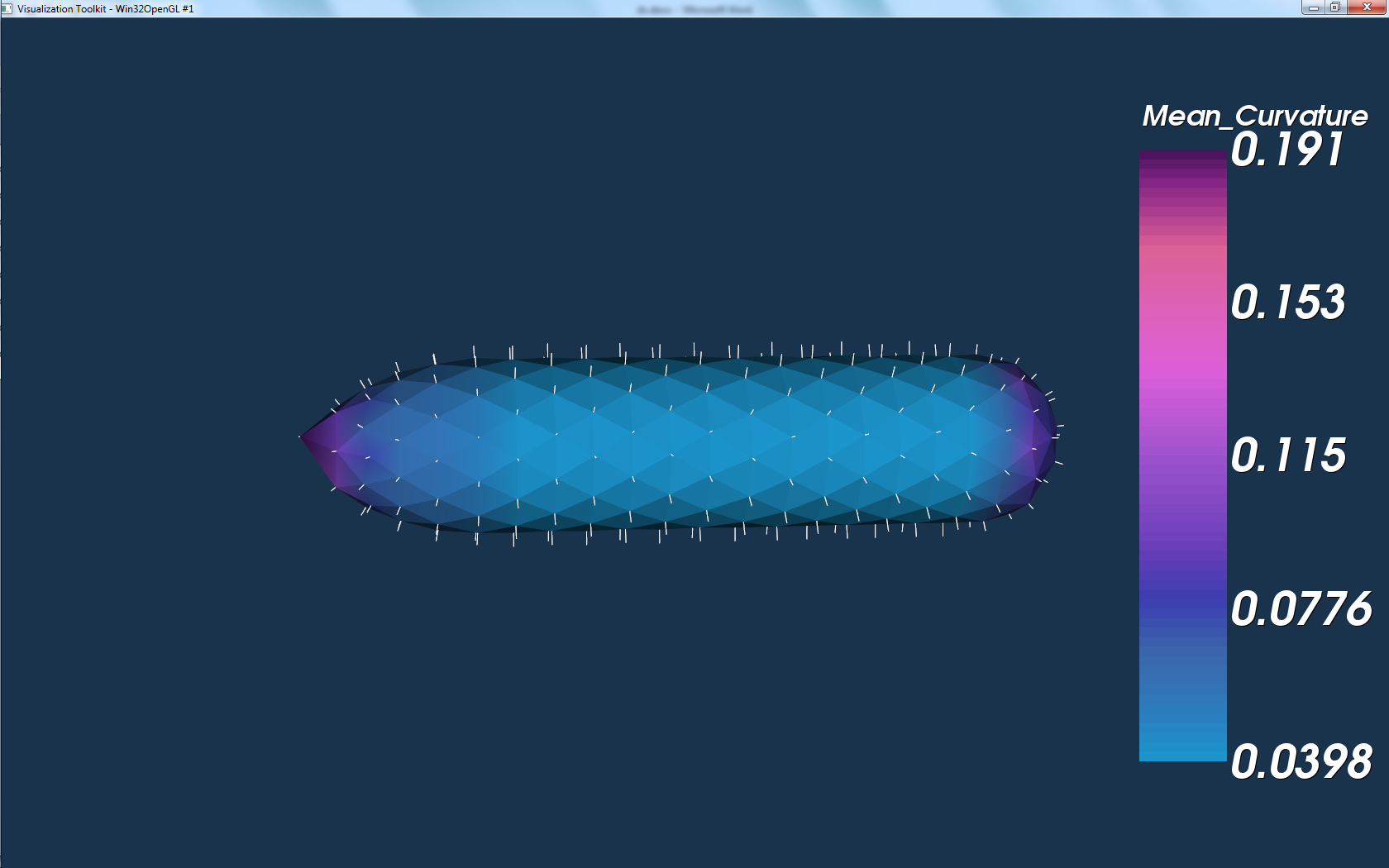}
  \end{minipage}
  \begin{minipage}{0.4\textwidth}
  \includegraphics[width=\textwidth]{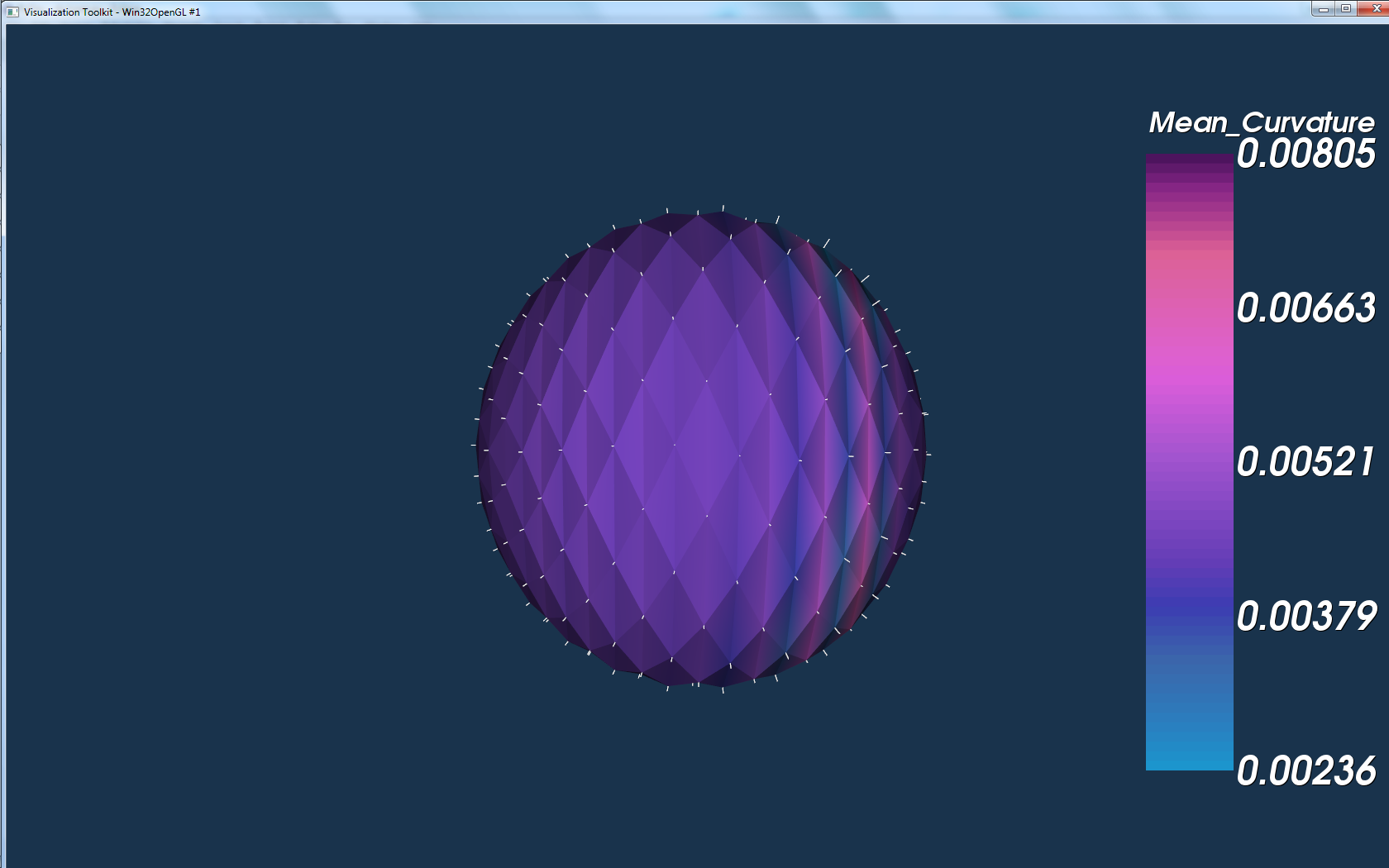}
  \end{minipage}
  \caption{Cylinder mesh initial (left) and result (right)}
  \label{fig12}
\end{figure}

\begin{figure}
  \centering
  \begin{minipage}{0.4\textwidth}
  \includegraphics[width=\textwidth]{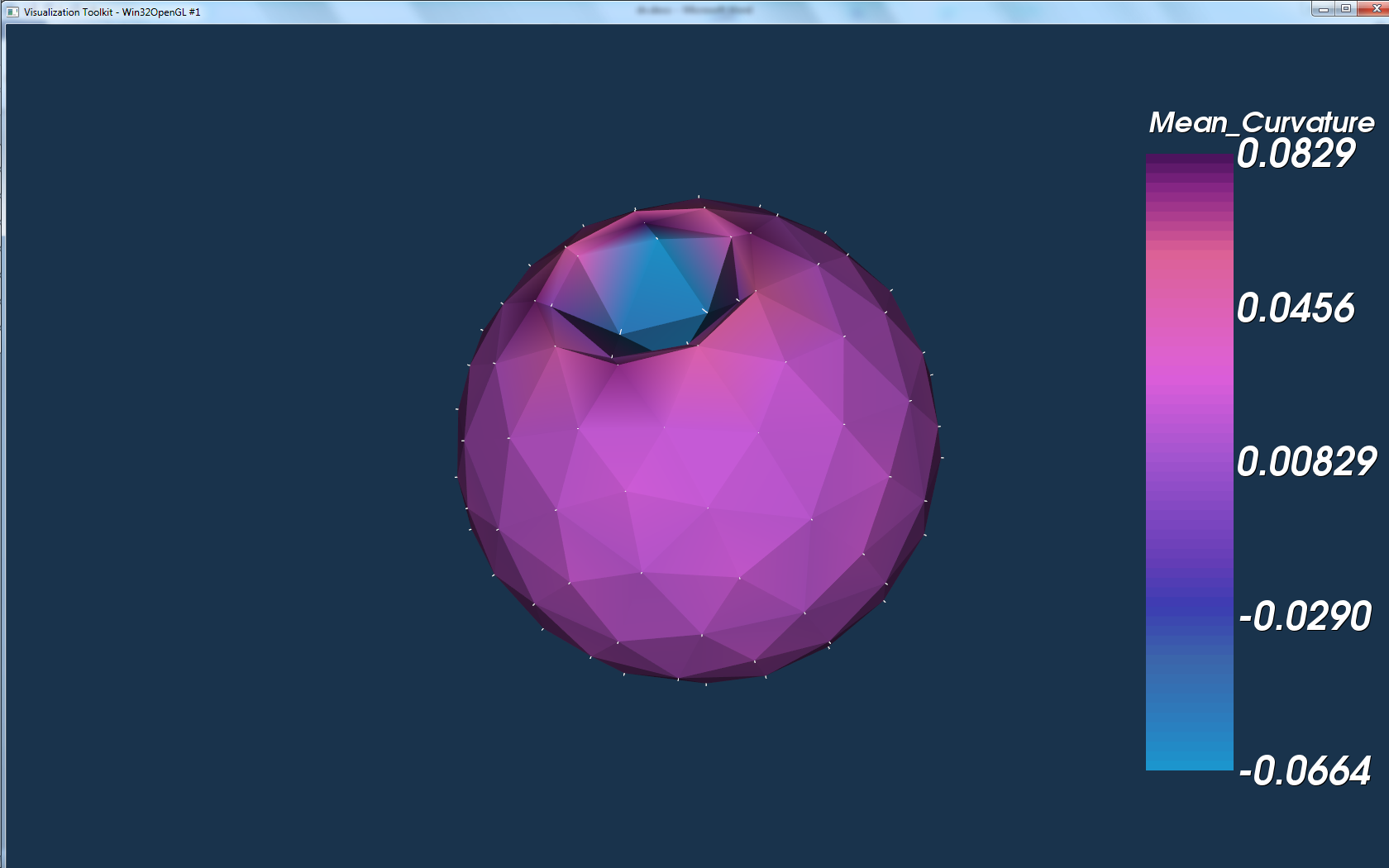}
  \end{minipage}
  \begin{minipage}{0.4\textwidth}
  \includegraphics[width=\textwidth]{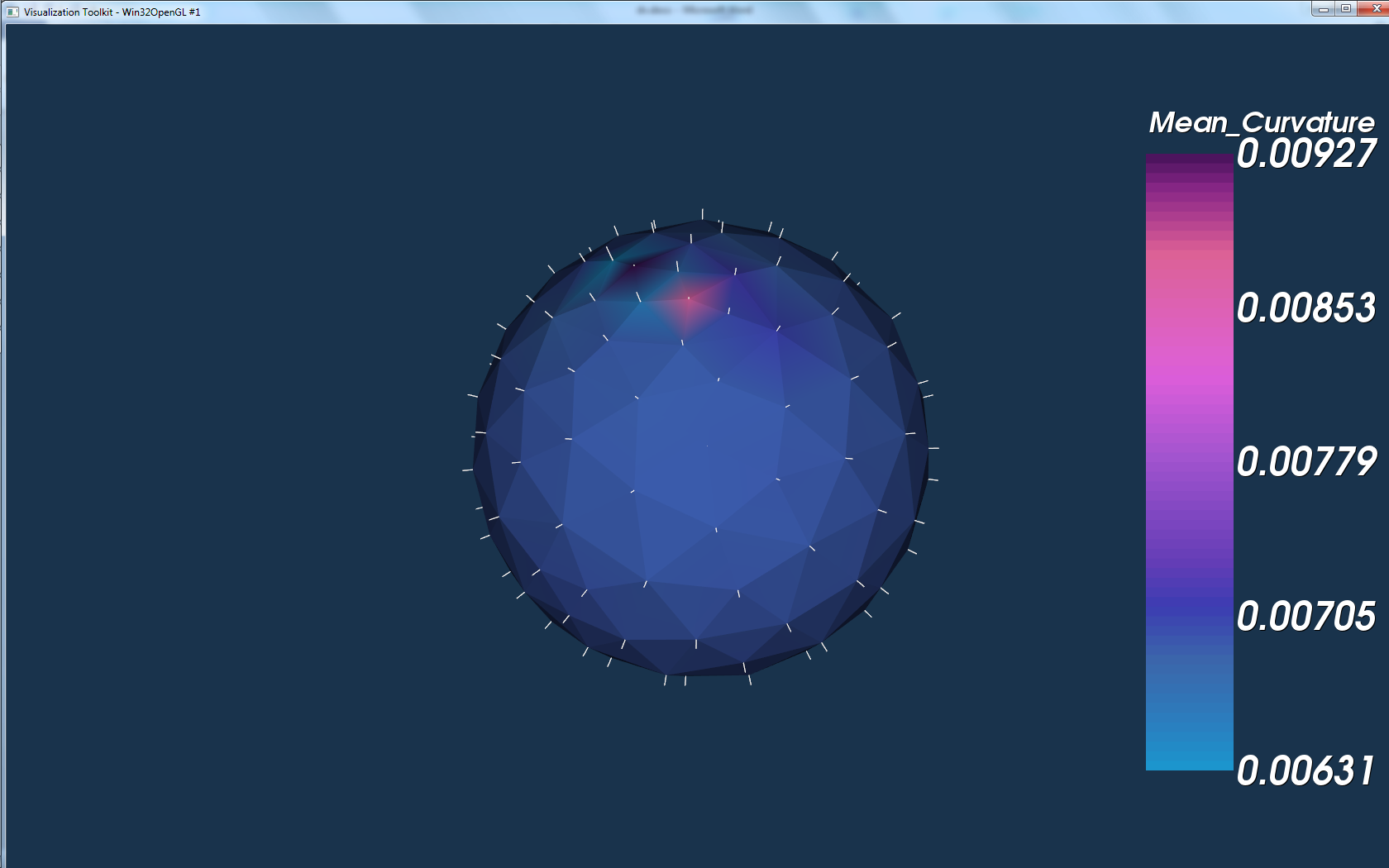}
  \end{minipage}
  \caption{Non-convex sphere mesh (left) and result (right)}
  \label{fig13}
\end{figure}

\clearpage
\bibliographystyle{alpha}
\bibliography{literature.bib}

\end{document}